\documentclass[10pt,prl,showpacs,twocolumn,nofootinbib,unsortedaddress,floatfix]{revtex4-1}
\usepackage{newtxtext,newtxmath}
\usepackage{bm}
\usepackage{physics}
\usepackage{mathtools,mathrsfs,amsfonts,dsfont}
\usepackage{ulem}
\usepackage[unicode=true, pdfusetitle, bookmarks=true, bookmarksnumbered=false, bookmarksopen=false, breaklinks=true, pdfborder={0 0 0}, backref=false, colorlinks=true, linkcolor=blue, citecolor=blue, urlcolor=blue]{hyperref}
\usepackage{cancel}
\usepackage[dvipsnames]{xcolor}
\newcommand{\sm}{\kern0.1em}

\begin{document}

\title{Comment on ``the Spin Dependence of Detection Times
and the Nonmeasurability of Arrival Times''}

\author{Siddhant Das}
\email{Siddhant.Das@physik.uni-muenchen.de}
\author{Serj Aristarhov} %put email
\email{S.Aristarkhov@campus.lmu.de}
\affiliation{Mathematisches Institut, Ludwig-Maximilians-Universit\"{a}t M\"{u}nchen, Theresienstr.\ 39, D-80333 M\"{u}nchen, Germany}
\date{Oct.\ 6, 2023}

\begin{abstract}
We respond to the recent article by S.\ Goldstein, R.\ Tumulka, and N.\ Zangh\`i [\href{https://arxiv.org/abs/2309.11835}{arXiv:2309.11835 [quant-ph]} (2023)] concerning the spin-dependent arrival-time distributions reported in [S. Das and D. D\"urr, \href{https://www.nature.com/articles/s41598-018-38261-4}{Sci.\ Rep.\ \textbf{9\,}: 2242} (2019)].
\end{abstract}

\maketitle
\normalem 
In a recent article \cite{GTZ}, Goldstein, Tumulka and Zangh\`i (GTZ in the sequel) claim that the arrival-time distributions of a spin-1/2 particle moving within a waveguide reported in \cite{DD,*Exotic} are ``not measurable'', i.e., it is impossible that the results of any actual experiment have the reported distributions. They base their conclusion on the following points:
\begin{enumerate}
    \item \label{I} \emph{Criterion of measurability:} The statistics of outcomes of any quantum experiment are governed by a Positive Operator-Valued Measure (POVM). The probability for the measured quantity to lie in some set $\smash{S\subset \mathbb{R}}$, in this case, takes the form \(\matrixel{\Psi_i}{O(S)}{\Psi_i}\), where \(O(S)\) is an experiment-specific positive operator and \(\ket{\Psi_i}\) is the prepared quantum state of the probed system. 
    
    \item \label{II} The arrival-time probability distributions reported in \cite{DD,*Exotic} contradict \ref{I}.
\end{enumerate}

In the arrival-time (or time-of-flight) experiment proposed in \cite{DD,*Exotic}, the particle is prepared in various space-spin separated (or spin-polarized) states \(\smash{\ket{\Psi_i}=\ket{\psi}\otimes\ket{\vu{n}}}\) with identical spatial-state \(\ket{\psi}\), but different spin states \(\ket{\vu{n}}\); \(\smash{\vu{n}\in\mathcal{S}^2}\) being the direction of spin-polarization. Let \(P_{\vu{n}}\) denote the probability distribution of the experimental outcomes when the particle is prepared with the spin-state \(\vu{n}\).

To demonstrate \ref{II}, GTZ show that a key property of spin-polarization-dependent probability distributions compatible with \ref{I}, \emph{viz.},
\begin{equation}\label{amazing}
    P_{\vu{n}} + P_{-\,\vu{n}} \sm\text{ is independent of } \vu{n}, 
\end{equation}
is violated by those reported in \cite{DD,*Exotic}, thus contradicting \ref{I}. We agree with this conclusion. Since the direction defined by the waveguide axis is distinguished in the setup of \cite{DD,*Exotic}, the failure to comply with \eqref{amazing} does not surprise us. However, we do not think that the criterion \ref{I} of measurability is suitable for judging with certainty what can or cannot be observed in actual experiments.

This criterion \cite[Sec.\ 7.1.]{DGZOperators}, GTZ assert, is a ``fundamental tenet in standard quantum mechanics,'' and in fact ``a theorem within the framework of Bohmian mechanics''---``an immediate consequence of the very meaning of an experiment from a Bohmian perspective'' \cite{DGZMeaning}. In what follows, we limit our discussion to Bohmian mechanics (BM hereafter) since this theory serves as the foundation for the theoretical calculations in \cite{DD,*Exotic}. At the same time, we note that the arrival-time problem \cite{MSP,*MUGA1} has long been challenging the ``all's POVM postulate of standard QM''. Among the many proposed solutions, some comply with it and some do not (e.g., \cite{Bondurant,*Maccone,*Pullin,PI1,*PI2,*PI3,TJN,Leon} that are \emph{non-bilinear} in the prepared state).

The ``all's POVM theorem in BM'' is proved in, e.g., \cite[Sec.\ 5.1.2.]{TumulkaBook}, \cite[Sec.\ 7.2]{DurrDustin}, \cite[Sec.\ 6]{DGZMeaning}, \cite{DGZOperators}. In order to effectively convey our objection, a brief explanation of the basic framing, core assumptions, and a sketch of the proof of this theorem are recalled next.

The proof is based on a Bohmian analysis of an experiment that takes place between times \(t_i\) and \(t_f\). The system of interest and the apparatus(es) it interacts with are combined into a single composite system, to which the wave function \(\smash{\Phi_i(x,y)=\Psi_i(x)\sm\Xi_i(y)}\) is assigned at time \(t_i\),\footnote{For instance, \(\Psi_i(x)\) would be the spin-polarized states introduced previously.} and the system (apparatus) particles are assumed to have the configuration \(X_i\) (\(Y_i\)). It is subsequently granted that
\begin{quote}
    the many-body wave function \(\Phi_i(x,y)\) evolves \emph{linearly} via Schr\"odinger's equation up until time \(t_f\).
\end{quote}
and, furthermore, that (quantum equilibrium hypothesis)
\begin{quote}
    the initial configuration of the composite system, \((X_i,Y_i)\), realized at the beginning of different runs of this experiment---each featuring the same composite wave function \(\Phi_i\)---is random with probability density \(|\Phi_i(x,y)|^2\).
\end{quote}
From these posits, it follows via a well-known consequence of the defining equations of BM that the system configuration at time \(t_f\), \((X_f,Y_f)\), is distributed with density \(|\Phi_f(x,y)|^2\). This enables one to calculate the statistics of the apparatus configuration \(Y_f\), which records the outcome of the experiment. From this point, a few lines of straightforward calculations establish that the statistics of the apparatus-configuration \(Y_f\), as a functional depending on the \emph{probed system's initial wave function} \(\Psi_i(x)\) must have the form dictated in \ref{I}. In particular, \emph{it must be bilinear in \(\Psi_i\)}, from which property \eqref{amazing} essentially follows.

While we do not doubt the correctness of the proof of the theorem, we do find one of its key underlying assumptions questionable. To explain why, we must clarify the meaning of the joint wave function \(\Phi(x,y)\) in BM.

In a universe ruled by BM, there is only one system that is a priori subject to it, namely the universe itself. However, experimental consequences of quantum theories, including BM, can only be deduced from subsystem wave functions, such as those of the macroscopic system discussed above or the single spin-1/2 particle used in \cite{DD,*Exotic}. In BM one can ascribe a so-called conditional wave function (CWF) to an \emph{arbitrary} subsystem of particles \cite[Sec.\ 2.2]{DGZOperators}, using the resources of the theory, \emph{viz.}, the universal wave function and the Bohmian trajectories of its constituent particles.\footnote{It should be noted that as yet it is unclear how to define CWFs for a spin-aware Bohmian theory. Taking this crucial issue into account would only strengthen our argument, but we will ignore it for the sake of simplicity.} Thus, \(\Phi_{i(f)}(x,y)\) encountered in the proof of the ``all's POVM theorem in BM'' must be understood---absent further hypotheses---as the CWF of the macroscopic system comprised of the particle and measuring apparatuses.

Once that is admitted, by its very construction, the CWF \emph{does not} evolve by Schr\"odinger's equation, as claimed in the proof of the ``all's POVM theorem in BM''. The equations governing the evolution of CWFs are \emph{highly non-linear and non-unitary} (see, e.g., \cite[Sec.\ 3.1, p.\ 3137]{Travis} for the CWF equation of one particle in a two-particle universe).\footnote{For this reason, in appropriate situations, Bohmian CWFs can explain the illusion of wave-function collapse.} The assumption of the linear evolution of $\Phi_i(x,y)$ is evidently crucial; without it the theorem does not hold.

It has been suggested that in certain special situations (e.g., where the effective wave function ansatz \cite[Eq.\ (2.7)]{DGZOperators} holds) the subsystem CWF actually behaves as an \emph{effective wave function}---one satisfying an autonomous Schr\"odinger evolution. But this constitutes a further assumption about the nature of the universal wave function. Of perhaps greater concern, this ansatz being \emph{preserved} over time between the instants \(t_i\) and \(t_f\), perhaps even longer, is more difficult to justify. 

At this point one could argue that if we doubt the effective wave function ansatz, we should also question ascribing to microscopic subsystems a proper wave function evolving according to the Schr\"odinger equation, as was done in~\cite{DD,*Exotic}. But the assumption that, after certain painstaking laboratory operations (which are usually called ``the state-preparation") and subsequent isolation, microscopic subsystems do possess a definite wave function evolving according to the Schr\"odinger equation, remains plausible even without theoretical justification. In fact, this is precisely the assumption made in the analysis of nearly every quantum experiment, leading to excellent accord between theory and observation. The same reasoning evidently does not apply to the case of macroscopic bodies like measuring equipment. In fact, we would expect that the CWF of macroscopic objects evolves non-linearly, accounting for their necessarily Newtonian behaviour; see \cite[Sec.\ V]{Bassi}, \cite{Oriols}.

To sum up, the proof of the ``all's POVM theorem in BM'' rests on the assumption that macroscopic systems (such as complete experimental setups) possess their own (effective) wave functions evolving according to the Schr\"odinger equation. In our view, this assumption is not plausible enough to use the consequences of the theorem to judge with certainty what can and cannot be seen in a real experiment.

The reader should not confuse our doubts with any sort of certainty that the assumption in question is wrong. We find it implausible for the reasons listed above, but by no means do we have proof of its incorrectness. It may be correct together with the conclusion of \cite{GTZ}, or it may be wrong and the empirical arrival-time distributions will resemble those calculated in \cite{DD,*Exotic}. The best way to make progress here is to perform the arrival-time experiment in question.

\bibliography{ref}

\begin{thebibliography}{20}%
\makeatletter
\providecommand \@ifxundefined [1]{%
 \@ifx{#1\undefined}
}%
\providecommand \@ifnum [1]{%
 \ifnum #1\expandafter \@firstoftwo
 \else \expandafter \@secondoftwo
 \fi
}%
\providecommand \@ifx [1]{%
 \ifx #1\expandafter \@firstoftwo
 \else \expandafter \@secondoftwo
 \fi
}%
\providecommand \natexlab [1]{#1}%
\providecommand \enquote  [1]{``#1''}%
\providecommand \bibnamefont  [1]{#1}%
\providecommand \bibfnamefont [1]{#1}%
\providecommand \citenamefont [1]{#1}%
\providecommand \href@noop [0]{\@secondoftwo}%
\providecommand \href [0]{\begingroup \@sanitize@url \@href}%
\providecommand \@href[1]{\@@startlink{#1}\@@href}%
\providecommand \@@href[1]{\endgroup#1\@@endlink}%
\providecommand \@sanitize@url [0]{\catcode `\\12\catcode `\$12\catcode
  `\&12\catcode `\#12\catcode `\^12\catcode `\_12\catcode `\%12\relax}%
\providecommand \@@startlink[1]{}%
\providecommand \@@endlink[0]{}%
\providecommand \url  [0]{\begingroup\@sanitize@url \@url }%
\providecommand \@url [1]{\endgroup\@href {#1}{\urlprefix }}%
\providecommand \urlprefix  [0]{URL }%
\providecommand \Eprint [0]{\href }%
\providecommand \doibase [0]{http://dx.doi.org/}%
\providecommand \selectlanguage [0]{\@gobble}%
\providecommand \bibinfo  [0]{\@secondoftwo}%
\providecommand \bibfield  [0]{\@secondoftwo}%
\providecommand \translation [1]{[#1]}%
\providecommand \BibitemOpen [0]{}%
\providecommand \bibitemStop [0]{}%
\providecommand \bibitemNoStop [0]{.\EOS\space}%
\providecommand \EOS [0]{\spacefactor3000\relax}%
\providecommand \BibitemShut  [1]{\csname bibitem#1\endcsname}%
\let\auto@bib@innerbib\@empty
%</preamble>
\bibitem [{\citenamefont {Goldstein}\ \emph {et~al.}(2023)\citenamefont
  {Goldstein}, \citenamefont {Tumulka},\ and\ \citenamefont
  {Zangh{\`{i}}}}]{GTZ}%
  \BibitemOpen
  \bibfield  {author} {\bibinfo {author} {\bibfnamefont {S.}~\bibnamefont
  {Goldstein}}, \bibinfo {author} {\bibfnamefont {R.}~\bibnamefont {Tumulka}},
  \ and\ \bibinfo {author} {\bibfnamefont {N.}~\bibnamefont {Zangh{\`{i}}}},\
  }\href@noop {} {\enquote {\bibinfo {title} {On the spin dependence of
  detection times and the nonmeasurability of arrival times},}\ } (\bibinfo
  {year} {2023}),\ \Eprint {http://arxiv.org/abs/2309.11835} {arXiv:2309.11835
  [quant-ph]} \BibitemShut {NoStop}%
\bibitem [{\citenamefont {Das}\ and\ \citenamefont {D{\"{u}}rr}(2019)}]{DD}%
  \BibitemOpen
  \bibfield  {author} {\bibinfo {author} {\bibfnamefont {S.}~\bibnamefont
  {Das}}\ and\ \bibinfo {author} {\bibfnamefont {D.}~\bibnamefont
  {D{\"{u}}rr}},\ }\href {\doibase 10.1038/s41598-018-38261-4} {\bibfield
  {journal} {\bibinfo  {journal} {Sci. Rep.}\ }\textbf {\bibinfo {volume}
  {9}},\ \bibinfo {pages} {2242} (\bibinfo {year} {2019})}\BibitemShut
  {NoStop}%
\bibitem [{\citenamefont {Das}\ \emph {et~al.}(2019)\citenamefont {Das},
  \citenamefont {N\"oth},\ and\ \citenamefont {D\"urr}}]{Exotic}%
  \BibitemOpen
  \bibfield  {author} {\bibinfo {author} {\bibfnamefont {S.}~\bibnamefont
  {Das}}, \bibinfo {author} {\bibfnamefont {M.}~\bibnamefont {N\"oth}}, \ and\
  \bibinfo {author} {\bibfnamefont {D.}~\bibnamefont {D\"urr}},\ }\href
  {\doibase 10.1103/PhysRevA.99.052124} {\bibfield  {journal} {\bibinfo
  {journal} {Phys. Rev. A}\ }\textbf {\bibinfo {volume} {99}},\ \bibinfo
  {pages} {052124} (\bibinfo {year} {2019})}\BibitemShut {NoStop}%
\bibitem [{\citenamefont {D{\"{u}}rr}\ \emph {et~al.}(2004)\citenamefont
  {D{\"{u}}rr}, \citenamefont {Goldstein},\ and\ \citenamefont
  {Zangh{\`{I}}}}]{DGZOperators}%
  \BibitemOpen
  \bibfield  {author} {\bibinfo {author} {\bibfnamefont {D.}~\bibnamefont
  {D{\"{u}}rr}}, \bibinfo {author} {\bibfnamefont {S.}~\bibnamefont
  {Goldstein}}, \ and\ \bibinfo {author} {\bibfnamefont {N.}~\bibnamefont
  {Zangh{\`{I}}}},\ }\href {\doibase
  https://doi.org/10.1023/B:JOSS.0000037234.80916.d0} {\bibfield  {journal}
  {\bibinfo  {journal} {J. Stat. Phys.}\ }\textbf {\bibinfo {volume} {116}},\
  \bibinfo {pages} {959} (\bibinfo {year} {2004})}\BibitemShut {NoStop}%
\bibitem [{\citenamefont {D{\"u}rr}\ \emph {et~al.}(1995)\citenamefont
  {D{\"u}rr}, \citenamefont {Goldstein},\ and\ \citenamefont
  {Zangh{\`{i}}}}]{DGZMeaning}%
  \BibitemOpen
  \bibfield  {author} {\bibinfo {author} {\bibfnamefont {D.}~\bibnamefont
  {D{\"u}rr}}, \bibinfo {author} {\bibfnamefont {S.}~\bibnamefont {Goldstein}},
  \ and\ \bibinfo {author} {\bibfnamefont {N.}~\bibnamefont {Zangh{\`{i}}}},\
  }\href@noop {} {\enquote {\bibinfo {title} {Bohmian mechanics and the meaning
  of the wave function},}\ } (\bibinfo {year} {1995}),\ \Eprint
  {http://arxiv.org/abs/quant-ph/9512031} {arXiv:quant-ph/9512031 [quant-ph]}
  \BibitemShut {NoStop}%
\bibitem [{\citenamefont {Muga}\ \emph {et~al.}(1998)\citenamefont {Muga},
  \citenamefont {Sala},\ and\ \citenamefont {Palao}}]{MSP}%
  \BibitemOpen
  \bibfield  {author} {\bibinfo {author} {\bibfnamefont {J.~G.}\ \bibnamefont
  {Muga}}, \bibinfo {author} {\bibfnamefont {R.}~\bibnamefont {Sala}}, \ and\
  \bibinfo {author} {\bibfnamefont {J.}~\bibnamefont {Palao}},\ }\href
  {\doibase https://doi.org/10.1006/spmi.1997.0544} {\bibfield  {journal}
  {\bibinfo  {journal} {Superlattices and Microstructures}\ }\textbf {\bibinfo
  {volume} {23}},\ \bibinfo {pages} {833 } (\bibinfo {year}
  {1998})}\BibitemShut {NoStop}%
\bibitem [{\citenamefont {Muga}\ and\ \citenamefont {Leavens}(2000)}]{MUGA1}%
  \BibitemOpen
  \bibfield  {author} {\bibinfo {author} {\bibfnamefont {J.~G.}\ \bibnamefont
  {Muga}}\ and\ \bibinfo {author} {\bibfnamefont {C.~R.}\ \bibnamefont
  {Leavens}},\ }\href {\doibase 10.1016/S0370-1573(00)00047-8} {\bibfield
  {journal} {\bibinfo  {journal} {Phys. Rep.}\ }\textbf {\bibinfo {volume}
  {338}},\ \bibinfo {pages} {353} (\bibinfo {year} {2000})}\BibitemShut
  {NoStop}%
\bibitem [{\citenamefont {Bondurant}(2004)}]{Bondurant}%
  \BibitemOpen
  \bibfield  {author} {\bibinfo {author} {\bibfnamefont {R.~S.}\ \bibnamefont
  {Bondurant}},\ }\href {\doibase 10.1103/PhysRevA.69.062104} {\bibfield
  {journal} {\bibinfo  {journal} {Phys. Rev. A}\ }\textbf {\bibinfo {volume}
  {69}},\ \bibinfo {pages} {062104} (\bibinfo {year} {2004})}\BibitemShut
  {NoStop}%
\bibitem [{\citenamefont {Maccone}\ and\ \citenamefont
  {Sacha}(2020)}]{Maccone}%
  \BibitemOpen
  \bibfield  {author} {\bibinfo {author} {\bibfnamefont {L.}~\bibnamefont
  {Maccone}}\ and\ \bibinfo {author} {\bibfnamefont {K.}~\bibnamefont
  {Sacha}},\ }\href {\doibase 10.1103/PhysRevLett.124.110402} {\bibfield
  {journal} {\bibinfo  {journal} {Phys. Rev. Lett.}\ }\textbf {\bibinfo
  {volume} {124}},\ \bibinfo {pages} {110402} (\bibinfo {year}
  {2020})}\BibitemShut {NoStop}%
\bibitem [{\citenamefont {Gambini}\ and\ \citenamefont
  {Pullin}(2022)}]{Pullin}%
  \BibitemOpen
  \bibfield  {author} {\bibinfo {author} {\bibfnamefont {R.}~\bibnamefont
  {Gambini}}\ and\ \bibinfo {author} {\bibfnamefont {J.}~\bibnamefont
  {Pullin}},\ }\href
  {http://iopscience.iop.org/article/10.1088/1367-2630/ac6768} {\bibfield
  {journal} {\bibinfo  {journal} {New J. Phys.}\ } (\bibinfo {year}
  {2022})}\BibitemShut {NoStop}%
\bibitem [{\citenamefont {Marchewka}\ and\ \citenamefont {Schuss}(1998)}]{PI1}%
  \BibitemOpen
  \bibfield  {author} {\bibinfo {author} {\bibfnamefont {A.}~\bibnamefont
  {Marchewka}}\ and\ \bibinfo {author} {\bibfnamefont {Z.}~\bibnamefont
  {Schuss}},\ }\href {\doibase 10.1016/S0375-9601(98)00107-8} {\bibfield
  {journal} {\bibinfo  {journal} {Phys. Lett. A}\ }\textbf {\bibinfo {volume}
  {240}},\ \bibinfo {pages} {177 } (\bibinfo {year} {1998})}\BibitemShut
  {NoStop}%
\bibitem [{\citenamefont {Marchewka}\ and\ \citenamefont {Schuss}(2001)}]{PI2}%
  \BibitemOpen
  \bibfield  {author} {\bibinfo {author} {\bibfnamefont {A.}~\bibnamefont
  {Marchewka}}\ and\ \bibinfo {author} {\bibfnamefont {Z.}~\bibnamefont
  {Schuss}},\ }\href {\doibase 10.1103/PhysRevA.63.032108} {\bibfield
  {journal} {\bibinfo  {journal} {Phys. Rev. A}\ }\textbf {\bibinfo {volume}
  {63}},\ \bibinfo {pages} {032108} (\bibinfo {year} {2001})}\BibitemShut
  {NoStop}%
\bibitem [{\citenamefont {Marchewka}\ and\ \citenamefont {Schuss}(2002)}]{PI3}%
  \BibitemOpen
  \bibfield  {author} {\bibinfo {author} {\bibfnamefont {A.}~\bibnamefont
  {Marchewka}}\ and\ \bibinfo {author} {\bibfnamefont {Z.}~\bibnamefont
  {Schuss}},\ }\href {\doibase 10.1103/PhysRevA.65.042112} {\bibfield
  {journal} {\bibinfo  {journal} {Phys. Rev. A}\ }\textbf {\bibinfo {volume}
  {65}},\ \bibinfo {pages} {042112} (\bibinfo {year} {2002})}\BibitemShut
  {NoStop}%
\bibitem [{\citenamefont {Juri{\'c}}\ and\ \citenamefont
  {Nikoli{\'c}}(2022)}]{TJN}%
  \BibitemOpen
  \bibfield  {author} {\bibinfo {author} {\bibfnamefont {T.}~\bibnamefont
  {Juri{\'c}}}\ and\ \bibinfo {author} {\bibfnamefont {H.}~\bibnamefont
  {Nikoli{\'c}}},\ }\href {https://doi.org/10.1140/epjp/s13360-022-02854-w}
  {\bibfield  {journal} {\bibinfo  {journal} {Eur. Phys. J. Plus}\ }\textbf
  {\bibinfo {volume} {137}} (\bibinfo {year} {2022})}\BibitemShut {NoStop}%
\bibitem [{\citenamefont {Le\'on}\ \emph {et~al.}(2000)\citenamefont {Le\'on},
  \citenamefont {Julve}, \citenamefont {Pitanga},\ and\ \citenamefont
  {de~Urr\'{\i}es}}]{Leon}%
  \BibitemOpen
  \bibfield  {author} {\bibinfo {author} {\bibfnamefont {J.}~\bibnamefont
  {Le\'on}}, \bibinfo {author} {\bibfnamefont {J.}~\bibnamefont {Julve}},
  \bibinfo {author} {\bibfnamefont {P.}~\bibnamefont {Pitanga}}, \ and\
  \bibinfo {author} {\bibfnamefont {F.~J.}\ \bibnamefont {de~Urr\'{\i}es}},\
  }\href {\doibase 10.1103/PhysRevA.61.062101} {\bibfield  {journal} {\bibinfo
  {journal} {Phys. Rev. A}\ }\textbf {\bibinfo {volume} {61}},\ \bibinfo
  {pages} {062101} (\bibinfo {year} {2000})}\BibitemShut {NoStop}%
\bibitem [{\citenamefont {Tumulka}(2022)}]{TumulkaBook}%
  \BibitemOpen
  \bibfield  {author} {\bibinfo {author} {\bibfnamefont {R.}~\bibnamefont
  {Tumulka}},\ }\enquote {\bibinfo {title} {General observables},}\ in\ \href
  {\doibase 10.1007/978-3-031-09548-1_5} {\emph {\bibinfo {booktitle}
  {Foundations of Quantum Mechanics}}}\ (\bibinfo  {publisher} {Springer
  International Publishing},\ \bibinfo {address} {Cham},\ \bibinfo {year}
  {2022})\ pp.\ \bibinfo {pages} {179--255}\BibitemShut {NoStop}%
\bibitem [{\citenamefont {D{\"{u}}rr}\ and\ \citenamefont
  {Lazarovici}(2020)}]{DurrDustin}%
  \BibitemOpen
  \bibfield  {author} {\bibinfo {author} {\bibfnamefont {D.}~\bibnamefont
  {D{\"{u}}rr}}\ and\ \bibinfo {author} {\bibfnamefont {D.}~\bibnamefont
  {Lazarovici}},\ }\href {\doibase https://doi.org/10.1007/978-3-030-40068-2}
  {\emph {\bibinfo {title} {Understanding Quantum Mechanics: The world
  according to modern quantum foundations}}}\ (\bibinfo  {publisher} {Springer
  International Publishing},\ \bibinfo {address} {Cham},\ \bibinfo {year}
  {2020})\BibitemShut {NoStop}%
\bibitem [{\citenamefont {Norsen}\ \emph {et~al.}(2015)\citenamefont {Norsen},
  \citenamefont {Marian},\ and\ \citenamefont {Oriols}}]{Travis}%
  \BibitemOpen
  \bibfield  {author} {\bibinfo {author} {\bibfnamefont {T.}~\bibnamefont
  {Norsen}}, \bibinfo {author} {\bibfnamefont {D.}~\bibnamefont {Marian}}, \
  and\ \bibinfo {author} {\bibfnamefont {X.}~\bibnamefont {Oriols}},\ }\href
  {\doibase https://doi.org/10.1007/s11229-014-0577-0} {\bibfield  {journal}
  {\bibinfo  {journal} {Synthese}\ }\textbf {\bibinfo {volume} {192}},\
  \bibinfo {pages} {3125} (\bibinfo {year} {2015})}\BibitemShut {NoStop}%
\bibitem [{\citenamefont {Toro{\v{s}}}\ \emph {et~al.}(2016)\citenamefont
  {Toro{\v{s}}}, \citenamefont {Donadi},\ and\ \citenamefont {Bassi}}]{Bassi}%
  \BibitemOpen
  \bibfield  {author} {\bibinfo {author} {\bibfnamefont {M.}~\bibnamefont
  {Toro{\v{s}}}}, \bibinfo {author} {\bibfnamefont {S.}~\bibnamefont {Donadi}},
  \ and\ \bibinfo {author} {\bibfnamefont {A.}~\bibnamefont {Bassi}},\ }\href
  {\doibase 10.1088/1751-8113/49/35/355302} {\bibfield  {journal} {\bibinfo
  {journal} {J. Phys. A: Math. Theor.}\ }\textbf {\bibinfo {volume} {49}},\
  \bibinfo {pages} {355302} (\bibinfo {year} {2016})}\BibitemShut {NoStop}%
\bibitem [{\citenamefont {Oriols}\ and\ \citenamefont
  {Benseny}(2017)}]{Oriols}%
  \BibitemOpen
  \bibfield  {author} {\bibinfo {author} {\bibfnamefont {X.}~\bibnamefont
  {Oriols}}\ and\ \bibinfo {author} {\bibfnamefont {A.}~\bibnamefont
  {Benseny}},\ }\href {\doibase 10.1088/1367-2630/aa719a} {\bibfield  {journal}
  {\bibinfo  {journal} {New Journal of Physics}\ }\textbf {\bibinfo {volume}
  {19}},\ \bibinfo {pages} {063031} (\bibinfo {year} {2017})}\BibitemShut
  {NoStop}%
\end{thebibliography}%
\end{document}